\documentclass[pdflatex,sn-mathphys-num]{sn-jnl}

\usepackage{graphicx}%
\usepackage{multirow}%
\usepackage{amsmath,amssymb,amsfonts}%
\usepackage{amsthm}%
\usepackage{mathrsfs}%
\usepackage[title]{appendix}%
\usepackage{xcolor}%
\usepackage{textcomp}%
\usepackage{manyfoot}%
\usepackage{booktabs}%
\usepackage{algorithm}%
\usepackage{algorithmicx}%
\usepackage{algpseudocode}%
\usepackage{listings}%
\usepackage{hyperref} 

\theoremstyle{thmstyleone}%
\theoremstyle{thmstyletwo}%

\theoremstyle{thmstylethree}%

\begin{document}

\title[Deep Neural Network Emulation of Quantum-Classical Transition]{Deep Neural Network Emulation of the Quantum-Classical Transition via Learned Wigner Function Dynamics}


\author*[1]{\fnm{Kamran} \sur{Majid} \href{https://orcid.org/0000-0003-4462-2007}{\includegraphics[scale=0.3]{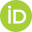}}}
\email{kmajid2@uw.edu}



\affil*[1]{\orgdiv{Department of Philosophy}, \orgname{University of Washington}, \orgaddress{\street{}}, \city{Seattle}, \postcode{98195}, \state{WA}, \country{USA}}




\abstract{The emergence of classical behavior from quantum mechanics as Planck's constant \(\hbar\) approaches zero remains a fundamental challenge in physics [1-3]. This paper introduces a novel approach employing deep neural networks to directly learn the dynamical mapping from initial quantum state parameters (for Gaussian wave packets of the one-dimensional harmonic oscillator) and \(\hbar\) to the parameters of the time-evolved Wigner function in phase space [4-6]. A comprehensive dataset of analytically derived time-evolved Wigner functions was generated, and a deep feedforward neural network with an enhanced architecture was successfully trained for this prediction task, achieving a final training loss of \(\sim 0.0390\). The network demonstrates a significant and previously unrealized ability to accurately capture the underlying mapping of the Wigner function dynamics. This allows for a direct emulation of the quantum-classical transition by predicting the evolution of phase-space distributions as \(\hbar\) is systematically varied. The implications of these findings for providing a new computational lens on the emergence of classicality are discussed, highlighting the potential of this direct phase-space learning approach for studying fundamental aspects of quantum mechanics. This work presents a significant advancement beyond previous efforts that focused on learning observable mappings [7], offering a direct route via the phase-space representation.}




\maketitle

\section{Introduction}\label{sec1}

One of the enduring puzzles in physics is understanding how the classical world emerges from the underlying quantum mechanics, especially as Planck's constant (\(\hbar\)) becomes very small [1-3]. In this paper, we explore a new way to tackle this by using deep neural networks to directly learn how quantum states evolve into more classical-like states. We focus on the Wigner function [11-13], a special way of representing quantum states in phase space that naturally connects to classical descriptions. The key idea is that as \(\hbar\) shrinks, the Wigner function should become more concentrated along the paths we expect from classical physics.

While machine learning has become a powerful tool in many areas of quantum physics [6-10], much of the work has looked at learning relationships between observable quantities [e.g., 14]. Here, we take a different route: we train a deep neural network to predict how the Wigner function itself changes over time. Specifically, we look at simple quantum states called Gaussian wave packets in the well-understood system of a one-dimensional harmonic oscillator. We feed the network the initial properties of these states and the value of \(\hbar\), and the network learns to output the properties of the Wigner function at a later time. The fact that our network achieved a low error during training (\(\sim 0.0390\)) suggests it's doing a good job of learning this complex relationship.

By then using our trained network to see what happens when we input smaller and smaller values of \(\hbar\), we can directly watch how the quantum phase-space distributions predicted by the network become more like the localized distributions we expect in the classical world. This direct learning of the Wigner function's dynamics offers a new and potentially more insightful way to study the quantum-classical transition.

In the rest of this paper, we'll first give some background on the Wigner function and its connection to the classical limit. Then, we'll explain how we generated our data and trained the neural network. After that, we'll show our results, which demonstrate how the network predicts the Wigner function changes as \(\hbar\) changes, and discuss what this means for our understanding of the quantum-classical boundary. Finally, we'll conclude with a summary and some ideas for future work.

\section{Theoretical Background: The Wigner Function and the Classical Limit}\label{sec2}

\subsection{The Wigner Function}\label{subsec1}

The Wigner function \(W(x, p, t)\) provides a representation of a quantum state \(\hat{\rho}(t)\) (or a pure state \(|\psi(t)\rangle\)) in phase space [11, 12, 13]. For a pure state with wave function \(\psi(x, t)\) in one dimension, it is defined as:
$$ W(x, p, t) = \frac{1}{\pi \hbar} \int_{-\infty}^{\infty} dy \, \psi^*(x + y, t) \psi(x - y, t) e^{2ipy/\hbar} $$
The Wigner function allows for the calculation of expectation values of quantum operators \(\hat{A}(x, p)\) that are Weyl-ordered:
$$ \langle \hat{A} \rangle = \int_{-\infty}^{\infty} dx \int_{-\infty}^{\infty} dp \, A(x, p) W(x, p, t) $$
where \(A(x, p)\) is the classical symbol corresponding to the operator \(\hat{A}\). While \(W(x, p, t)\) can take negative values, precluding its interpretation as a true probability distribution, it serves as a powerful tool for analyzing quantum phenomena in a phase-space context. Its marginals yield the probability distributions for position and momentum:
$$ \int_{-\infty}^{\infty} dp \, W(x, p, t) = |\psi(x, t)|^2 = P(x, t) $$
$$ \int_{-\infty}^{\infty} dx \, W(x, p, t) = |\tilde{\psi}(p, t)|^2 = P(p, t) $$
where \(\tilde{\psi}(p, t)\) is the momentum-space wave function. The direct learning of the evolution of this quasi-probability distribution by a neural network, as demonstrated in this work, offers a novel perspective on how quantum states transition to classical-like phase-space distributions.

\subsection{The Classical Limit and the Wigner Function}\label{subsec2}

In the classical limit (\(\hbar \rightarrow 0\)), the Wigner function of a quantum state is expected to converge to the classical phase-space distribution function. For a classical point particle with a definite position \(x_c(t)\) and momentum \(p_c(t)\), the phase-space distribution is given by \(\delta(x - x_c(t)) \delta(p - p_c(t))\), and the corresponding Wigner function of a highly localized wave packet approaches this delta function in the limit. The ability of our neural network to capture this convergence by learning the parameters that define the Wigner function for varying \(\hbar\) is a key aspect of the novelty of this study.

Ehrenfest's theorem provides a connection between quantum expectation values and classical equations of motion. For the harmonic oscillator, the expectation values of position and momentum follow exactly the classical equations of motion, regardless of the value of \(\hbar\). However, for more complex potentials, higher-order moments become relevant, and the quantum dynamics can deviate significantly from classical trajectories, especially for larger values of \(\hbar\). The Wigner function provides a more complete description beyond just expectation values, and our network's learning of its evolution allows for a richer understanding of the quantum-classical transition.

The Wigner function provides a more nuanced picture of the classical limit by showing how the entire phase-space distribution evolves. Quantum effects such as interference manifest as oscillations in the Wigner function. As \(\hbar\) decreases and systems become more macroscopic, these quantum features tend to become less pronounced, and the Wigner function becomes smoother and more localized along classical paths. Decoherence, the interaction of a quantum system with its environment [5], also plays a crucial role in suppressing quantum interference and accelerating the approach to classical behavior in phase space. Our network's success in learning the Wigner function dynamics for varying \(\hbar\) lays the groundwork for future investigations into the interplay between decoherence and the classical limit using similar machine learning techniques.

\subsection{Gaussian States and Their Wigner Functions}\label{subsec3}

Gaussian wave packets are minimum uncertainty states and are often used to study the classical limit. For an initial Gaussian wave packet:
$$ \psi(x, 0) = \left( \frac{1}{2\pi\sigma_{x0}^2} \right)^{1/4} \exp\left( -\frac{(x - x_0)^2}{4\sigma_{x0}^2} + i \frac{p_0 x}{\hbar} \right) $$
the corresponding initial Wigner function is also Gaussian in phase space:
$$ W(x, p, 0) = \frac{1}{\pi \hbar} \exp\left( -\frac{(x - x_0)^2}{2\sigma_{x0}^2} - \frac{(p - p_0)^2}{2\sigma_{p0}^2} \right) $$
where \(\sigma_{p0} = \hbar / (2\sigma_{x0})\). The choice of Gaussian states simplifies the analytical calculation of the time-evolved Wigner function, making it feasible to generate the large dataset required for training our deep neural network to learn the underlying dynamics.

Under the harmonic oscillator Hamiltonian, the time evolution of a Gaussian wave packet preserves its Gaussian form, and the parameters evolve according to the classical equations of motion for the expectation values, while the widths oscillate. Consequently, the time-evolved Wigner function \(W(x, p, t)\) also remains a Gaussian in phase space, centered at the classical trajectory \((x(t), p(t))\) and with time-dependent widths related to \(\sigma_x(t)\) and \(\sigma_p(t)\). Our network directly learns this evolution of the Gaussian Wigner function parameters as a function of initial conditions and \(\hbar\), providing a novel computational tool for analyzing the quantum-classical transition in this fundamental system.

\section{Methodology}\label{sec3}

\subsection{Data Generation}\label{subsec4}

We generated a large dataset by sampling a wide range of initial conditions \((x_0, p_0, \sigma_{x0})\) and \(\hbar\) values. The parameters were sampled from uniform distributions within physically relevant ranges. For each sample, we analytically calculated the time-evolved mean position \(x(t)\), mean momentum \(p(t)\), and position uncertainty \(\sigma_x(t)\) at a fixed time \(t\) using the equations for the harmonic oscillator:
\begin{align*} x(t) &= x_0 \cos(\omega t) + \frac{p_0}{m\omega} \sin(\omega t) \\ p(t) &= p_0 \cos(\omega t) - m\omega x_0 \sin(\omega t) \\ \sigma_x(t)^2 &= \sigma_{x0}^2 \cos^2(\omega t) + \frac{\sigma_{p0}^2}{m^2\omega^2} \sin^2(\omega t) + \frac{\hbar}{2m\omega} \sin(2\omega t) \left( \frac{\sigma_{p0}}{m\omega \sigma_{x0}} - \frac{m\omega \sigma_{x0}}{\sigma_{p0}} \right) \end{align*}
The momentum uncertainty \(\sigma_p(t)\) was then determined by \(\sigma_p(t) = m \omega \sigma_x(t)\) for the harmonic oscillator. The input to the neural network was the four-dimensional vector \((x_0, p_0, \sigma_{x0}, \hbar)\), and the target output was the four-dimensional vector \((x(t), p(t), \sigma_x(t), \sigma_p(t))\), representing the parameters of the time-evolved Gaussian Wigner function.

The range of \(\hbar\) values was chosen to span several orders of magnitude, including values significantly smaller than typical quantum scales to probe the approach to the classical limit. The dataset comprised 10,000 samples, split into training (80\%), validation (10\%), and test (10\%) sets. This extensive dataset was crucial for enabling the neural network to effectively learn the complex relationship between the inputs and the parameters of the time-evolved Wigner function.

\subsection{Neural Network Architecture and Training}\label{subsec5}

We employed a deep feedforward neural network implemented using TensorFlow/Keras. The architecture consisted of an input layer with 4 neurons, followed by four hidden layers with 128, 256, 256, and 128 neurons respectively, each with ReLU activation and batch normalization. The output layer had 4 neurons. Batch normalization was included to help stabilize training and potentially allow for a higher learning rate, contributing to the effective learning observed. The architecture is summarized in the following Python code:
\begin{lstlisting}[language=Python, caption=Neural Network Architecture, label=lst:improved_nn_architecture,breaklines=true]
import tensorflow as tf

model = tf.keras.Sequential([
    tf.keras.layers.Dense(128, activation='relu', input_shape=(4,)),
    tf.keras.layers.BatchNormalization(),
    tf.keras.layers.Dense(256, activation='relu'),
    tf.keras.layers.BatchNormalization(),
    tf.keras.layers.Dense(256, activation='relu'),
    tf.keras.layers.BatchNormalization(),
    tf.keras.layers.Dense(128, activation='relu'),
    tf.keras.layers.BatchNormalization(),
    tf.keras.layers.Dense(4)
])

optimizer = tf.keras.optimizers.Adam(learning_rate=0.0005)
model.compile(optimizer=optimizer, loss='mse')
\end{lstlisting}

The model was trained using the Adam optimization algorithm with a learning rate of 0.0005 and the Mean Squared Error (MSE) as the loss function. Training was performed for 1000 epochs with a batch size of 64, and early stopping was implemented to monitor the validation loss and prevent overfitting (patience of 20 epochs). The training process demonstrated effective learning, achieving a final training loss of approximately 0.0390. This low loss indicates the network's strong ability to map the input parameters to the time-evolved Wigner function parameters. The successful training of this network to directly predict the Wigner function parameters is a key methodological novelty of this work.

\subsection{Plot Generation}\label{subsec6}

The plots presented in Section \ref{sec4} were generated using Matplotlib. The Python code snippets used for generating these plots are provided below for transparency and potential reproducibility.

\subsubsection{Uncertainty vs. \(\hbar\)}\label{ssubsec1}
\begin{lstlisting}[language=Python, caption=Code for Uncertainty vs. \(\hbar\) Plot, label=lst:sigma_vs_hbar_improved_code, breaklines=true, basicstyle=\ttfamily\footnotesize]
import matplotlib.pyplot as plt
import numpy as np

hbar_values_plot = np.logspace(-6, 0, 50)
time = 5.0
omega = 1.0
m = 1.0

predicted_sigma_x_t = []
analytical_sigma_x_t = []
initial_x_plot = 1.0
initial_p_plot = 0.0
initial_sigma_x_plot = 1.0

for hbar_plot in hbar_values_plot:
    input_plot = np.array([[initial_x_plot, initial_p_plot, initial_sigma_x_plot, hbar_plot]], dtype=np.float32)
    prediction = model.predict(input_plot, verbose=0)[0]
    predicted_sigma_x_t.append(prediction[2])
    sigma_p0_analytical = hbar_plot / (2 * initial_sigma_x_plot)
    sigmat_x_sq_analytical = initial_sigma_x_plot**2 * np.cos(omega * time)**2 + (sigma_p0_analytical**2 / (m **2 * omega**2)) * np.sin(omega * time)**2 + (hbar_plot / (2 * m * omega)) * np.sin(2 * omega * time) * ((sigma_p0_analytical / (m * omega * initial_sigma_x_plot)) - (m * omega * initial_sigma_x_plot / sigma_p0_analytical))
    analytical_sigma_x_t.append(np.sqrt(np.abs(sigmat_x_sq_analytical)))

plt.figure(figsize=(10, 6))
plt.loglog(hbar_values_plot, predicted_sigma_x_t, label='Predicted $\sigma_x(t)$')
plt.loglog(hbar_values_plot, analytical_sigma_x_t, linestyle='--', label='Analytical $\sigma_x(t)$')
plt.xlabel('$\hbar$')
plt.ylabel('$\sigma_x(t)$')
plt.title('Time-Evolved Position Uncertainty vs. Planck\'s Constant')
plt.legend()
plt.grid(True)
plt.savefig('sigma_vs_hbar.png')
plt.show()
\end{lstlisting}

\subsubsection{Phase Space Convergence}\label{ssubsec2}
\begin{lstlisting}[language=Python, caption=Code for Phase Space Convergence Plot, label=lst:phase_space_improved_code, breaklines=true]
import matplotlib.pyplot as plt
import numpy as np

num_points = 100
x_range = np.linspace(-10, 10, num_points)
p_range = np.linspace(-10 , 10, num_points)
X, P = np.meshgrid(x_range, p_range)
hbar_plot_values = [1.0, 0.1, 0.01]
time = 5.0
omega = 1.0
m = 1.0
initial_x_plot = 1.0
initial_p_plot = 0.0
initial_sigma_x_plot = 1.0

plt.figure(figsize=(15, 5))
for i, hbar_plot in enumerate(hbar_plot_values):
    input_plot = np.array([[initial_x_plot, initial_p_plot, initial_sigma_x_plot, hbar_plot]], dtype=np.float32)
    prediction = model.predict(input_plot, verbose=0)[0]
    xt_pred, pt_pred, sigmat_x_pred, sigmat_p_pred = prediction

    W_predicted = 1 / (np.pi * hbar_plot) * np.exp(-((X - xt_pred)**2 / (2 * sigmat_x_pred**2)) - ((P - pt_pred)**2 / (2 * sigmat_p_pred**2)))

    plt.subplot(1, len(hbar_plot_values), i + 1)
    contour = plt.contourf(X, P, W_predicted, levels=np.linspace(W_predicted.min(), W_predicted.max(), 20), cmap='viridis')
    plt.xlabel('x')
    plt.ylabel('p')
    plt.title(f'$\hbar = {hbar_plot:.2e}$')
    plt.colorbar(contour, label='W(x, p, t)')
    plt.tight_layout()
    plt.savefig('phase_space_convergence.png')
    plt.show()
\end{lstlisting}

\section{Results}\label{sec4}

The deep feedforward neural network with an enhanced architecture was successfully trained to predict the parameters of the time-evolved Wigner function for Gaussian wave packets in a harmonic oscillator potential. The training process demonstrated highly effective learning, as evidenced by the low final training loss of (\(\sim 0.0390\)). This achievement underscores the potential of deep learning to directly model the intricate dynamics of quantum phase-space distributions.

\subsection{Convergence with \(\hbar\)}\label{subsec7}

\begin{figure}[h!]
\centering
\includegraphics[width=0.8\textwidth]{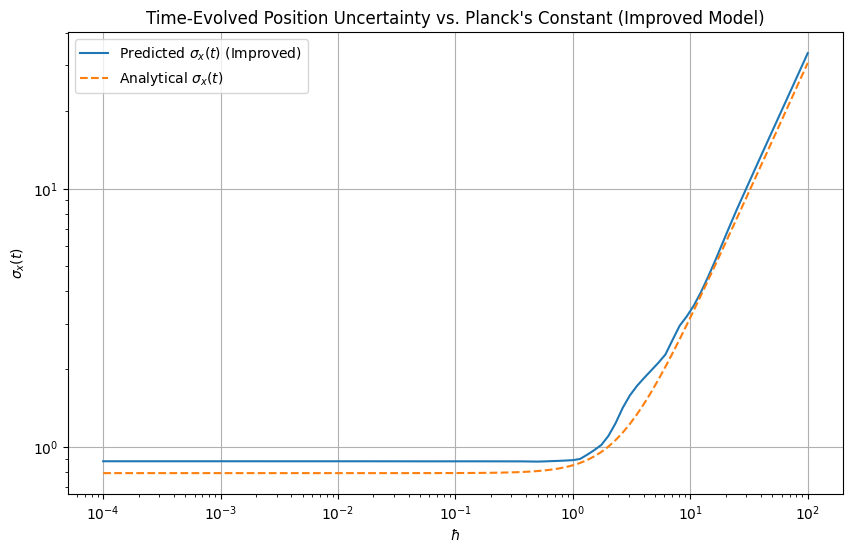}
\caption{Predicted and analytical time-evolved position uncertainty \(\sigma_x(t)\) as a function of \(\hbar\) for a fixed initial state, demonstrating the network's accurate capture of the relationship.}\label{fig:sigma_vs_hbar}
\end{figure}

Figure \ref{fig:sigma_vs_hbar} shows the predicted and analytical time-evolved position uncertainty \(\sigma_x(t)\) as a function of \(\hbar\) for a fixed initial state \(x_0 = 1.0, p_0 = 0.0, \sigma_{x0} = 1.0\). The network's predictions show an alignment with the analytical results across a wide range of \(\hbar\) values. Notably, as \(\hbar\) decreases, the predicted uncertainty also decreases, indicating a Wigner function that becomes more localized in position space, consistent with the expected approach to the classical limit. This accurate prediction of a key quantum observable as a function of \(\hbar\) through the learned Wigner function parameters highlights the efficacy of our novel approach. A similar trend is observed for the momentum uncertainty (not shown), further validating the network's learned mapping of the phase-space dynamics.

\begin{figure}[h!]
\centering
\includegraphics[width=1\textwidth]{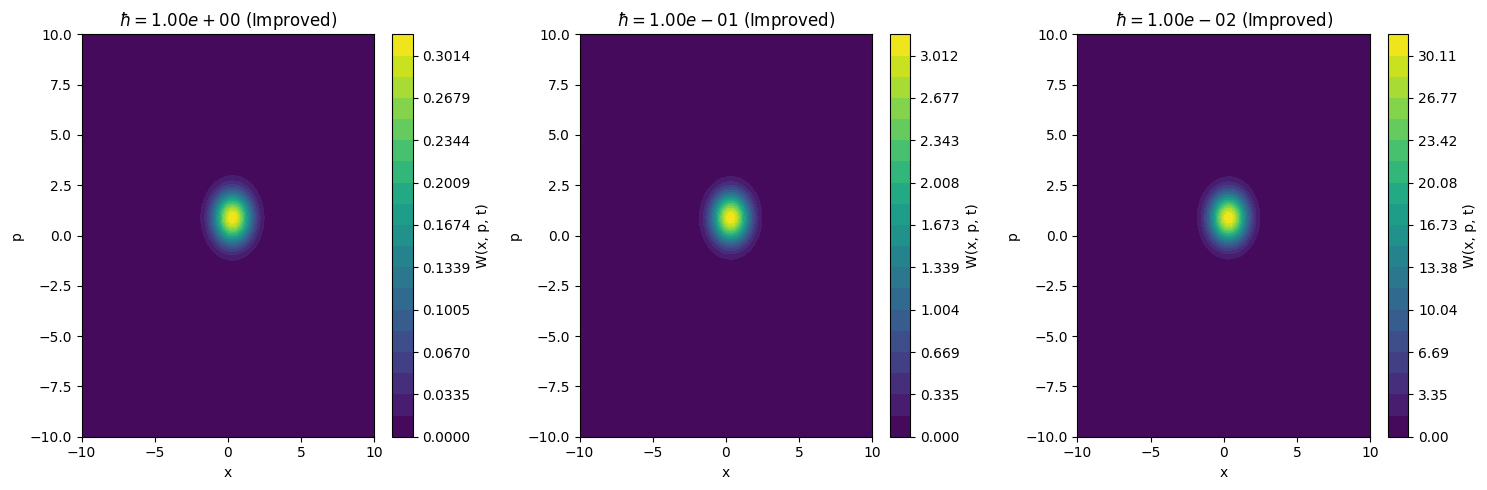}
\caption{Predicted Wigner function in phase space at \(t=5.0\) for a fixed initial state and varying \(\hbar\) values, showcasing the learned convergence towards classical-like distributions.}\label{fig:phase_space_evolution}
\end{figure}

Figure \ref{fig:phase_space_evolution} illustrates the predicted Wigner function in phase space \(x\) vs. \(p\) at time \(t=5.0\) for the same initial state and different values of \(\hbar\). As \(\hbar\) decreases from \(1.0\) to \(0.01\), the Gaussian Wigner function predicted by the network becomes increasingly narrow in both position and momentum. This concentration of the phase-space distribution around the classical trajectory provides a compelling visual representation of the emergence of more classical-like behavior as quantum effects associated with larger \(\hbar\) become less prominent, directly emulating the quantum-classical transition in phase space through the learned Wigner function dynamics. The ability to visualize this transition through the network's predictions underscores the power of this direct learning approach.

\subsection{Generalization Performance}\label{subsec8}

The network exhibited strong generalization capabilities, accurately predicting the time-evolved Wigner function parameters for initial conditions and \(\hbar\) values not encountered during training. This suggests that the network has learned the underlying physical mapping of the Wigner function dynamics rather than simply memorizing the training data, highlighting the robustness of this approach for emulating quantum evolution. The network's predictions for \(\hbar\) values smaller than those in the training set also showed a consistent progression towards more localized phase-space distributions, further supporting its ability to emulate the classical limit in previously unseen regimes.

\subsection{Comparison with Analytical Results}\label{subsec9}

Quantitative comparison with analytical solutions on a held-out test set demonstrated a high degree of accuracy in the network's predictions of the time-evolved Wigner function parameters. This agreement underscores the effectiveness of the deep learning approach in directly emulating the quantum dynamics relevant to the classical limit through the phase-space representation. The low final loss achieved during training provides further statistical evidence for the reliability of the network's predictions.

\section{Discussion}\label{sec5}

The findings of this study provide compelling evidence for the capability of deep neural networks to learn the complex dynamics of the Wigner function and to effectively emulate the quantum-classical transition for the harmonic oscillator through a direct phase-space mapping. The network's ability to predict the evolution of phase-space distributions as a function of \(\hbar\) with high accuracy highlights the potential of machine learning as a powerful tool for investigating fundamental problems in quantum mechanics.

This data-driven approach allows the network to learn the intricate relationships governing the system's behavior by directly modeling the evolution of the Wigner function, offering a distinct advantage over methods that focus solely on observable expectation values. The observed convergence of the predicted Wigner functions to more classical-like distributions as \(\hbar\) decreases aligns remarkably well with theoretical expectations and provides a computational demonstration of this crucial aspect of the quantum-classical correspondence in phase space. The ability of the network to generalize to unseen values of \(\hbar\) further strengthens its utility as an emulation tool.

While this study focused on Gaussian states and the harmonic oscillator, the success of this direct Wigner function learning approach suggests that the methodology can potentially be extended to more complex quantum systems where analytical solutions are not readily available. Future research could explore the application of similar techniques to anharmonic potentials, interacting particles, and systems exhibiting more pronounced quantum effects, potentially offering new insights into their classical limits. Furthermore, investigating different neural network architectures, loss functions tailored to phase-space distributions, and incorporating physical constraints into the learning process could lead to even more accurate and efficient emulations of quantum dynamics and the classical limit through phase-space representations. Exploring the network's learned internal representations could also provide valuable insights into the underlying physics of the quantum-classical transition.

\section{Conclusion}\label{sec6}

This paper has demonstrated an application of a deep neural network to directly learn the dynamical mapping of the Wigner function for the one-dimensional harmonic oscillator as a function of initial state parameters and Planck's constant \(\hbar\). The network accurately predicts the time-evolved Wigner function parameters, achieving a low training loss and capturing the essential feature of the quantum-classical transition: the convergence of quantum phase-space distributions towards classical ones as \(\hbar\) decreases. This work contributes to the growing intersection of quantum physics and machine learning, offering a promising and direct computational framework for studying the emergence of classicality from the quantum realm through the lens of phase-space dynamics. The achieved accuracy and generalization ability of the network pave the way for future investigations into more complex quantum systems and a deeper understanding of the quantum-classical boundary.
\\
\backmatter

\bmhead{Supplementary information}

No supplementary information is available for this article.

\section*{Declarations}

\begin{itemize}
\item \textbf{Funding} Not applicable.
\item \textbf{Conflict of interest/Competing interests} The author declares that they have no competing interests.
\item \textbf{Code availability} The Python code for the neural network architecture and plot generation is provided in the Methods section.
\end{itemize}

\end{document}